\documentclass[acmsmall,screen]{acmart}

\AtBeginDocument{%
  \providecommand\BibTeX{{%
    \normalfont B\kern-0.5em{\scshape i\kern-0.25em b}\kern-0.8em\TeX}}}

\usepackage{tcolorbox}
\usepackage{pgfplotstable}
\definecolor{linecolor}{RGB}{105,105,105} 
\definecolor{bgcolor}{RGB}{234,234,234} 
\usepackage{caption}
\usepackage{multirow}
\usepackage{booktabs}
\usepackage{booktabs}
\usepackage{tabularx}
\usepackage{array}
\usepackage{xcolor}
\usepackage{pifont}
\newcommand{\cmark}{\textcolor{green!60!black}{\ding{51}}}
\newcommand{\xmark}{\textcolor{red!80!black}{\ding{55}}}
\usepackage{multirow}
\usepackage{enumitem}

\usepackage[utf8]{inputenc}
\usepackage[T1]{fontenc}
\usepackage{algorithmic}
\usepackage{graphicx}
\usepackage{textcomp}
\usepackage{xcolor}
\usepackage{multirow, graphics}
\usepackage{listings}
\usepackage{diagbox}
\usepackage{subfig}
\usepackage{balance}
\usepackage{adjustbox}
\usepackage{colortbl}
\usepackage{framed}
\usepackage{booktabs}
\usepackage{indentfirst}
\usepackage{makecell,tcolorbox}
\usepackage[framemethod=TikZ]{mdframed}
\newmdenv[
  linecolor=linecolor,
  leftline=true,
  topline=false,
  bottomline=false,
  rightline=false,
  linewidth=2pt,
  innerleftmargin=10pt,
  innerrightmargin=10pt,
  innertopmargin=5pt,
  innerbottommargin=5pt,
  backgroundcolor=bgcolor
]{mycallout}
\usepackage[linesnumbered,ruled,vlined]{algorithm2e}
\usepackage{tcolorbox}
\tcbuselibrary{breakable}
\usepackage{tikz}
\usepackage{pgfplots}
\usepackage{subcaption}
\pgfplotsset{compat=1.18}
\usepgfplotslibrary{groupplots}
\usepackage{pgfplots}
\usepgfplotslibrary{groupplots}
\pgfplotsset{compat=1.18}

\definecolor{sodagrey}{RGB}{220,220,220}
\definecolor{sodagreen}{RGB}{190,225,180}
\definecolor{sodayellow}{RGB}{245,205,80}
\definecolor{sodared}{RGB}{210,130,130}
\definecolor{sodablue}{RGB}{150,175,215}

\newcommand{\benchname}{\textsc{ReqClarifyBench}}

\author{Mingyu Chen}
\email{mingyuchen1@link.cuhk.edu.cn}
\affiliation{
  \institution{The Chinese University of Hong Kong, Shenzhen}
  \state{Guangdong}
  \country{China}
}

\author{Zhiqing Zhong}
\email{zhiqingzhong@link.cuhk.edu.cn}
\authornotemark[1]
\affiliation{
  \institution{The Chinese University of Hong Kong, Shenzhen}
  \state{Guangdong}
  \country{China}
}

\author{Manyi Wang}
\email{manyiwang@link.cuhk.edu.cn}
\affiliation{
  \institution{The Chinese University of Hong Kong, Shenzhen}
  \state{Guangdong}
  \country{China}
}

\author{Linxi Liang}
\email{linxiliang@link.cuhk.edu.cn}
\affiliation{
  \institution{The Chinese University of Hong Kong, Shenzhen}
  \state{Guangdong}
  \country{China}
}

\author{Rouyu Chen}
\email{rouyuchen@link.cuhk.edu.cn}
\affiliation{
  \institution{The Chinese University of Hong Kong, Shenzhen}
  \state{Guangdong}
  \country{China}
}

\author{Pinjia He}
\email{hepinjia@cuhk.edu.cn}
\authornote{Corresponding author.}
\affiliation{
  \institution{The Chinese University of Hong Kong, Shenzhen}
  \state{Guangdong}
  \country{China}
}

\def\renderinsertedpart{}

\ifx\renderinsertedpart\undefined

\newenvironment{insertedpartenv}{}{}
\else

\fi

\begin{document}

\title{Don't Guess, Just Ask: Evaluating Interactive Requirement Clarification for Repository-Level Software Development}

\begin{abstract}
  With the rapid advancement of Large Language Models (LLMs), software development is gradually moving toward a paradigm where users describe what they want and LLMs carry out the implementation. As LLMs’ implementation capabilities improve, an increasingly critical bottleneck emerges: whether LLMs truly understand what users want. In practice, users often begin with vague intent, leaving many important requirements to be progressively clarified through interaction. This is especially consequential in repository-level development, where an incorrect LLM's guess about user intent can propagate through subsequent development phases, causing costly rework in long-horizon tasks.
However, existing repository-level code generation benchmarks typically evaluate code generation directly from either complete specifications or underspecified requests, overlooking the importance of pre-coding requirement clarification. 
Meanwhile, prior requirement clarification studies mainly focus on function-level tasks, failing to reflect the complexity of real-world software development scenarios. 
Thus, it remains unclear how well LLMs can turn vague user intent into clear requirements through interactive clarification and how such clarification capability affects subsequent code generation.

To bridge this gap, we introduce \benchname{}, an interactive requirement clarification benchmark for repository-level software development. \benchname{} contains 341 tasks spanning game, miniapp, and web development. Each task includes a high-level vague user intent, a detailed requirement document validated by human, and the corresponding code evaluation artifacts. To better simulate a realistic interactive clarification process, we further employ LLM-based user simulators with diverse personas derived from real-world conversation history.
Based on \benchname{}, we conduct a systematic empirical study. Specifically, we evaluate six LLMs’ requirement clarification capabilities from two complementary perspectives: how effectively they ask clarification questions and the quality of the generated requirement documents. 
The experimental results demonstrate that stronger clarification effectiveness does not necessarily translate into higher requirement document quality, and that LLMs exhibit substantial performance variations across different user personas. 
We also find that, compared with standalone LLMs, coding-agent harnesses do not further improve the ability to transform vague user intent into requirement documents.
Based on these findings, we provide valuable insights for building LLM-based end-to-end software development systems.

\end{abstract}



\newcommand{\figref}[1]{Figure~\ref{#1}}
\newcommand{\tabref}[1]{Table~\ref{#1}}
\newcommand{\secref}[1]{Section~\ref{#1}}

\newcommand{\toolName}[1]{\textbf{TraceDev\raisebox{-0.7ex}{\Huge\textbf{#1}}}}

\newcommand{\toolNameplain}[1]{TraceDev\raisebox{-0.7ex}{\Huge #1}}

\newcommand{\toolNameSmall}{TraceDev}
\newcommand{\datasetetour}{ETOUR}
\newcommand{\datasetsmos}{SMOS}
\newcommand{\ourgithub}{\url{https://github.com/ISSE-Lab/ISSTA2026-TraceDev}}
\newcommand{\ourdoi}{\url{https://zenodo.org/records/20809458}}

\maketitle
\newcommand{\revise}[1]{\textcolor{blue}{[revise: #1]}}

\newcommand{\ie}{\mbox{\emph{i. e.,\ }}}

\section{Introduction}\label{sec:introduction}
Large Language Models (LLMs) are becoming increasingly capable of handling long-horizon software development tasks~\cite{deng2026swemilestone,deng2026swebench}.
This progress is paving the way for a requirement-driven development paradigm, where users describe what they want and LLMs carry out the long-horizon implementation~\cite{tracedev,requirementareallyouneed,Requirementstocode,requirementdevelopmentase}. 
However, the success of this paradigm depends on clear and complete requirements, while real-world users often begin with vague descriptions of their intents that need to be progressively clarified through further interactions~\cite{ferrari2022requirements,vijayvargiya2026ambigswe}.
Neglecting requirement clarification may lead to incorrect assumptions, which will propagate to subsequent design, implementation, and testing phases, causing the implementation to drift away from the user's intent and increasing the need for human oversight and rework in long-horizon tasks~\cite{Impact,frattini2023requirements}.

\begin{figure*}[t]
    \centering
    \includegraphics[width=\textwidth]{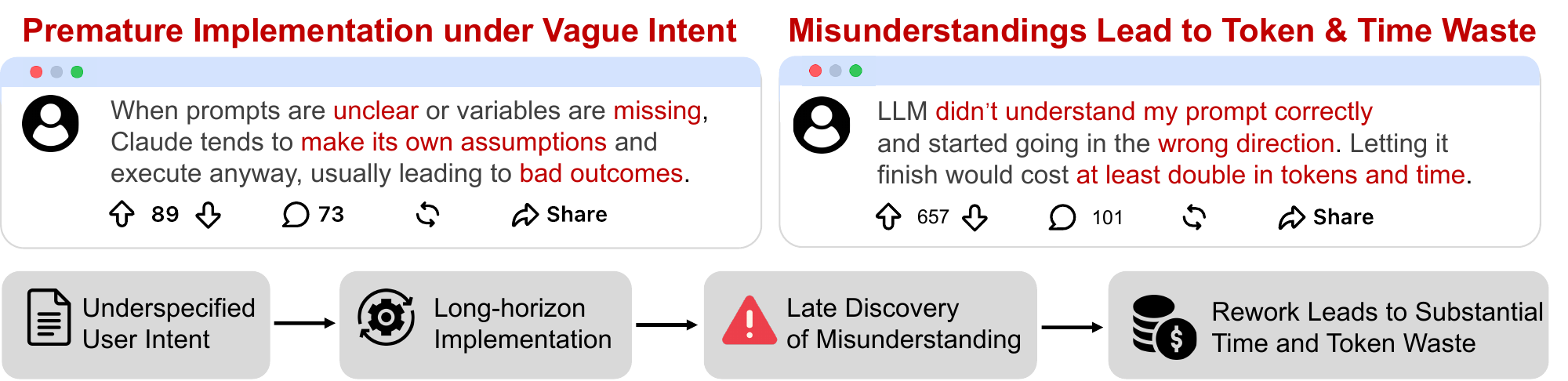}
    \caption{One complaint describes incorrect assumptions leading to an undesirable implementation; the other describes time and tokens wasted as the model continues along a misunderstood path.}
    \label{fig:intro-clarification}
    \vspace{-2mm}
\end{figure*}

Figure~\ref{fig:intro-clarification} presents two representative real-world user complaints that illustrate a shared failure mode. When a request is underspecified, LLMs may make a guess about what the user wants without asking for clarification, wasting time and tokens without producing a satisfactory outcome. Although LLMs’ implementation capabilities continue to improve, as reflected in rising benchmark scores~\cite{deng2026swebench,deng2026swemilestone,deepseekv41flash,claudeopus55}, they may still spend substantial time and tokens acting on a misunderstanding of the user’s request. Turning vague user intents into clear requirements through interactive clarification is therefore becoming an increasingly critical bottleneck.

Existing benchmarks~\cite{zhang2026miniappbench,luo2026gamecraftbenchagentsbuildplayable,lu2026webgenbench,ding2026nlrepobench,zhou2026featurebench} have demonstrated progress in the end-to-end transformation of Natural Language Requirements (NLRs) into repository-level code. However, they largely overlook the importance of requirement clarification in this process, evaluating code generation from either well-specified requirement documents or vague user requests without multi-turn requirement clarification.
Meanwhile, the Requirements Engineering (RE) community has explored LLM-based requirement clarification. Nevertheless, existing studies~\cite{ClarifyCodeBench,mu2024clarifygpt,yang2026assessing} mainly focus on function-level tasks, failing to reflect the complexity of real-world software projects. 
Moreover, these studies often employ simulated users that follow a fixed interaction pattern, failing to capture diverse real-user behaviors~\cite{jin2026reqelicitgym}.


To address these limitations, we introduce \textbf{\benchname{}}, an interactive requirement clarification benchmark for repository-level software development. \benchname{} aims to evaluate LLMs’ ability to transform vague user intents into clear requirement documents through multi-turn requirement clarification. \benchname{} contains 341 tasks spanning game, miniapp, and web development. It is constructed through a semi-automated pipeline.
Each task includes a high-level vague user intent, a detailed requirement document validated by humans, and the corresponding code evaluation artifacts. To support realistic interactions, we derive three representative user personas from SWE-chat~\cite{baumann2026swechatcodingagentinteractions}, a large-scale dataset of
multi-turn conversations between software developers and coding agents. These personas are used to guide the LLM-based user simulator throughout the clarification process. In addition, we introduce three new metrics to assess LLMs’ requirement clarification capabilities from two perspectives: how effectively LLMs ask clarification questions and the quality of the generated requirement documents.

Based on \benchname{}, we conduct a systematic empirical study focusing on the following four dimensions. \textbf{\textit{(1) Overall performance across LLMs and user personas.}} The results show that stronger clarification effectiveness does not necessarily lead to higher requirement document quality, and that user interaction styles can substantially affect model performance. \textbf{\textit{(2) LLMs’ clarification behavior.}} The results reveal three distinct interaction patterns: clarification with expert users is typically short and efficient, interactions with vague users are sustained but low-yield, and new requirement information often emerges in later rounds when users change their requirements. \textbf{\textit{(3) The impact of clarification on requirement document quality and code generation.}} Our experiments show that interactive clarification consistently improves requirement document quality, with particularly substantial gains in fine-grained requirement alignment, and further benefits subsequent implementation. 
\textbf{\textit{(4) Effect of coding-agent harness.}} We compare standalone LLMs with the same LLMs running within coding-agent harnesses. The results show that coding-agent harnesses improve clarification effectiveness but do not further enhance the quality of derived requirement documents.
Overall, these findings suggest that requirement clarification goes beyond simply asking more questions. Requirement clarification requires the evaluated LLM to elicit key information, understand and integrate requirements expressed during multi-turn interaction, and adapt its clarification strategy to different user personas. These capabilities together determine whether an LLM can transform a vague user intent into reliable software requirements, supporting subsequent code generation.

The main contributions of our work are summarized as follows:
\begin{itemize}
    \item \textbf{\benchname{}:} We introduce \benchname{}, the first interactive requirement clarification benchmark for repository-level software development, aiming to evaluate LLMs’ ability to transform vague user intents into clear and complete requirement documents through multi-turn interaction.
 In addition, we design three new evaluation metrics
to assess LLMs’ interactive requirement clarification capabilities.
    \item \textbf{Realistic User Simulation:} We construct three representative user personas based on a large-scale dataset of
multi-turn conversations between software developers and coding agents, aiming to support realistic multi-turn requirement clarification.
    \item \textbf{Empirical Study:} We conduct an empirical study to evaluate the requirement clarification capabilities of different LLMs and their impact on subsequent code generation, providing a systematic understanding of both current model capabilities and the role of requirement clarification in LLM-based software development.
    \item \textbf{Data Availability:} We open-source the code
and data at \cite{reqclarifybench2026}.
\end{itemize}

\section{Background \& Related Work}\label{sec:related}
\subsection{LLM-based Repository-level Code Generation}


LLM-based code generation has advanced from function-level programming to end-to-end repository-level software engineering~\cite{chen2021evaluating,mbpp,jimenez2024swebench,yang2024neurips-sweagent,hong2024metagpt,qian2024chatdev,luo2025rpg}.
Several approaches are proposed to facilitate this progress. MetaGPT~\cite{hong2024metagpt} employs agents with different roles to collaboratively perform requirement analysis, system design, and code implementation. ChatDev~\cite{qian2024chatdev} coordinates role-playing agents through communication across the design, coding, and testing stages. RPG~\cite{luo2025rpg} uses a structured Repository Planning Graph to support planning and repository generation. However, these approaches typically treat the initial user intent as a fixed task objective. Although requirement analysis is included, missing information is generally inferred by LLMs rather than clarified through interaction with the user. Consequently, the resulting software may deviate from the user’s actual intent. The conversational capabilities of LLMs enable them to proactively ask clarification questions and construct a clear and complete requirement document based on user feedback. Nevertheless, interactive requirement clarification remains underexplored in repository-level software development. In this paper, we introduce \benchname{} and conduct an empirical study to investigate the role of requirement clarification in repository-level software development.

\begin{table*}[t]
\centering

\caption{
Comparison of \benchname{} with representative code generation benchmarks in terms of publication year, task granularity, dataset size, requirement characteristics, and support for interactive requirement clarification. ``R.'' denotes ``Requirement'' and Character. denotes ``characteristic''.
}

\label{tab:benchmark_comparison}

\small
\setlength{\tabcolsep}{5.5pt}
\renewcommand{\arraystretch}{1.08}

\rowcolors{2}{white}{gray!10}

\begin{tabularx}{\textwidth}{
    l
    c
    l
    r
    >{\raggedright\arraybackslash}X
    c
}
\toprule

\textbf{Benchmark}
& \textbf{Time}
& \textbf{Granularity}
& \textbf{\#Tasks}
& \textbf{R. Character.}
& \textbf{R. Clarification} \\

\midrule

HumanEval~\cite{chen2021evaluating}
& 2021
& Function-level
& 164
& Clear\&Complete
& \xmark \\

MBPP~\cite{mbpp}
& 2021
& Function-level
& 974
& Clear\&Complete
& \xmark \\

SWE-bench~\cite{jimenez2024swebench}
& 2023
& Pull Request-level
& 2,294
& Clear\&Complete
& \xmark \\

LiveCodeBench~\cite{jain2025livecodebench}
& 2024
& Function-level
& 400+
& Clear\&Complete
& \xmark \\

WebGen-Bench~\cite{lu2026webgenbench}
& 2025
& Repository-level
& 101
& Underspecified
& \xmark \\

FeatureBench~\cite{zhou2026featurebench}
& 2026
& Repository-level
& 200
& Clear\&Complete
& \xmark \\

NL2RepoBench~\cite{ding2026nlrepobench}
& 2026
& Repository-level
& 104
& Clear\&Complete
& \xmark \\

MiniAppBench~\cite{zhang2026miniappbench}
& 2026
& Repository-level
& 500
& Underspecified
& \xmark \\

GameCraft-Bench~\cite{luo2026gamecraftbenchagentsbuildplayable}
& 2026
& Repository-level
& 140
& Clear\&Complete
& \xmark \\

ClarifyCodeBench~\cite{ClarifyCodeBench}
& 2026
& Function-level
& 419
& Underspecified
& \cmark \\
\midrule
\textbf{\benchname{}}
& \textbf{2026}
& \textbf{Repository-level}
& \textbf{341}
& \textbf{Underspecified}
& \textbf{\cmark} \\

\bottomrule
\end{tabularx}
\end{table*}

\subsection{Code Generation Benchmarks} 

 As shown in Table ~\ref{tab:benchmark_comparison}, early code generation benchmarks primarily focus on function-level tasks. HumanEval~\cite{chen2021evaluating}, MBPP~\cite{mbpp}, and LiveCodeBench~\cite{jain2025livecodebench} require LLMs to generate correct programs or functions from natural language problem descriptions. SWE-bench~\cite{jimenez2024swebench} subsequently extends evaluation to real-world code repositories. Its tasks are derived from GitHub issues and their corresponding pull requests, requiring LLMs to understand existing repositories and generate code modifications that resolve the reported issues. Recent benchmarks have further shifted toward end-to-end repository-level software development. FeatureBench~\cite{zhou2026featurebench} evaluates the implementation of complex features in existing repositories. WebGen-Bench~\cite{lu2026webgenbench} and NL2Repo-Bench~\cite{ding2026nlrepobench} assess the generation of multi-file websites and complete code repositories from natural language requirements, respectively. MiniAppBench~\cite{zhang2026miniappbench} and GameCraft-Bench~\cite{luo2026gamecraftbenchagentsbuildplayable} further extend evaluation to the end-to-end generation of applications and playable games.


Despite the growing task complexity, these benchmarks typically evaluate code generation from fixed requirements. In real-world development, however, users often begin with a vague intent, leaving many important requirements to be progressively clarified through interactions. 
Benchmarks that provide complete and clear descriptions~\cite{chen2021evaluating,mbpp,jain2025livecodebench,jimenez2024swebench,zhou2026featurebench,ding2026nlrepobench,luo2026gamecraftbenchagentsbuildplayable} overlook the requirement clarification process, failing to reflect real-world development scenarios.
Conversely, benchmarks using vague requests as inputs~\cite{lu2026webgenbench,zhang2026miniappbench} require LLMs to infer missing information without user feedback, making it difficult to determine whether failures stem from limited code generation capabilities or requirement misunderstandings. This highlights the need to systematically evaluate LLMs' ability to transform vague and incomplete user intents into clear requirement documents through interactions.




\subsection{Interactive Requirement Clarification}
Requirement clarification is a central problem in Requirements Engineering (RE)~\cite{1663532,yang2026assessing}. It typically involves iterative interaction with stakeholders to clarify their goals, constraints, and unstated requirements~\cite{skatehoder}. 
Recent studies have explored requirement clarification for LLM-based code generation from different perspectives. ClarifyGPT~\cite{mu2024clarifygpt} detects ambiguous requirements, generates targeted clarification questions, and refines the requirements based on user responses before generating code. ClarifyCoder~\cite{wu2025codelanguagemodelslearn} fine-tunes LLMs to recognize underspecified problems and request missing information. ClarifyCodeBench~\cite{ClarifyCodeBench} introduces an interactive benchmark containing underspecified programming problems, annotated clarification questions and answers, and executable test cases. However, these studies primarily focus on simple function-level problems, failing to reflect real-world complex software development.
ReqElicitGym~\cite{jin2026reqelicitgym} models requirement clarification as a multi-turn elicitation process, but its simulated user is modeled as a single oracle user following predefined interaction rules, which fails to capture the diversity of interaction styles and behavioral patterns among real users. Moreover, ReqElicitGym does not further investigate the impact of requirement clarification on subsequent code generation.
In contrast, our proposed \benchname{} evaluates interactive requirement clarification for repository-level software development using LLM-based user simulators with diverse personas.

\section{\benchname}\label{sec:ftdroid}
\begin{figure}
    \centering
    \includegraphics[width=1\linewidth]{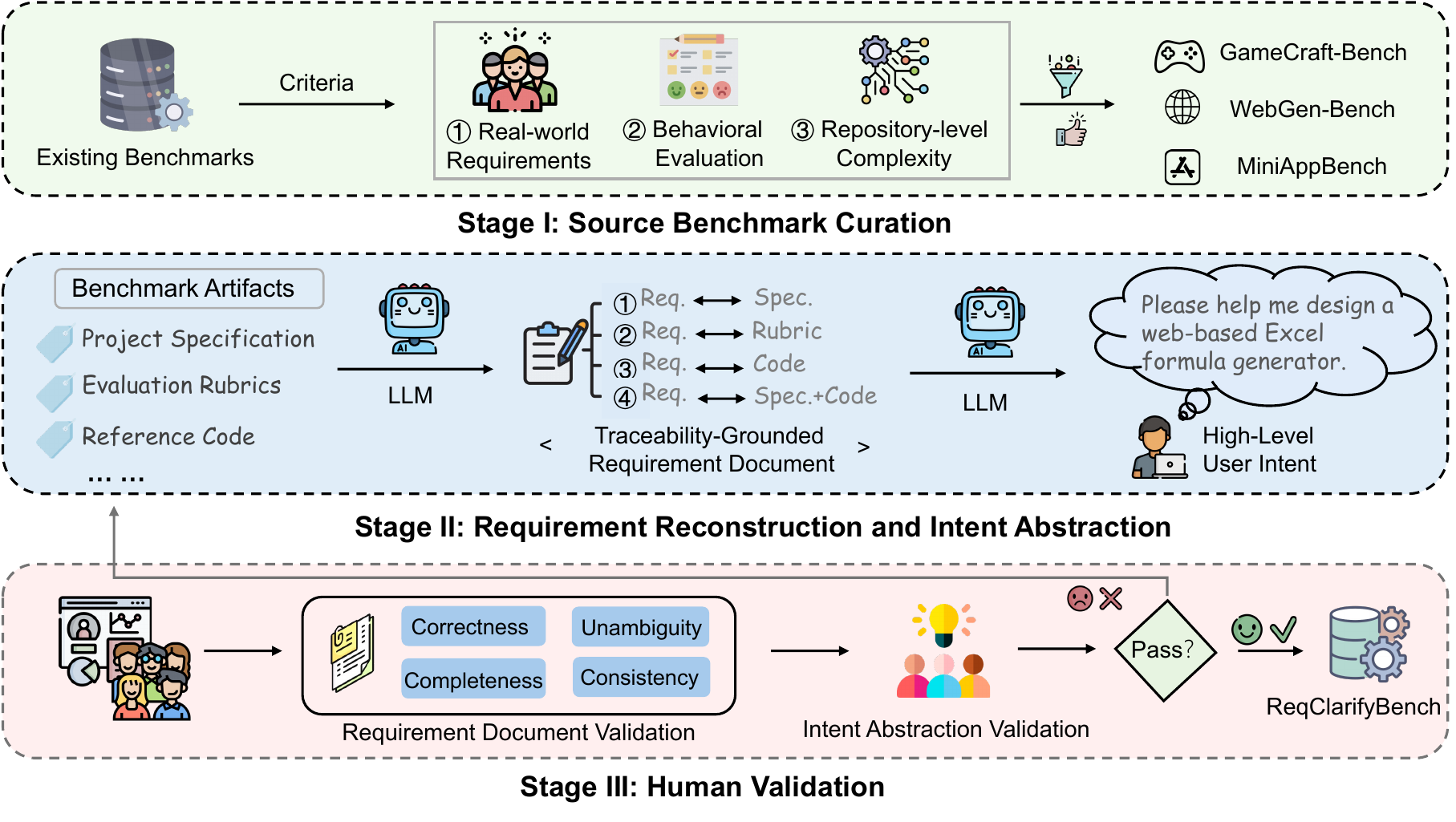}
    \caption{The overview of \benchname{} construction pipeline.}
    \label{fig:overview}
    \vspace{-2mm}
\end{figure}

In this section, we introduce \benchname{}, the first interactive requirement clarification benchmark for repository-level software development, aiming to evaluate LLMs’ ability to transform vague user intents into clear requirement documents through multi-turn interaction. We first describe the construction procedure and then present the characteristics of \benchname{}.

\subsection{Benchmark Construction}
As illustrated in Figure ~\ref{fig:overview}, \benchname{} is constructed through a semi-automated human-in-the-loop pipeline, which consists of three stages: source benchmark curation, requirement reconstruction and intent abstraction, and human validation. The pipeline is designed to enable adaptation to new datasets. Each task instance ultimately contains three core components: a high-level vague user intent, a complete and clear requirement document, and the corresponding code evaluation artifacts for assessing the generated code.

\textbf{Stage I: Source benchmark curation.}
Following~\cite{yang2026assessing,yang2025swebench,qiu2025how,han-etal-2024-archcode}, we construct \benchname{} based on existing code generation benchmarks rather than designing tasks from scratch. Existing benchmarks provide high-quality software development scenarios and reliable code evaluation protocols. Building on existing benchmarks allows us to focus on data construction and evaluation design for the requirement clarification stage. Meanwhile, we can reuse their code evaluation protocols to measure the impact of requirement clarification on subsequent implementation.

For source Benchmark Curation, we consider three main criteria. \textbf{\textit{1) Real-world Requirements.}} The tasks and development scenarios should be grounded in real-world requirements, rather than artificially constructed programming problems. \textbf{\textit{2) Repository-level Complexity.}} The tasks should target repository-level code generation rather than simple function completion, reflecting the scale and complexity of real-world software development. \textbf{\textit{3) Behavioral Evaluation.}} The benchmark should provide reliable behavioral evaluation. We avoid evaluations that heavily rely on unit tests with rigid interface assumptions, as they may penalize functionally correct implementations that do not conform to predefined APIs or code structures.

\begin{figure}
    \centering
    \includegraphics[width=1\linewidth]{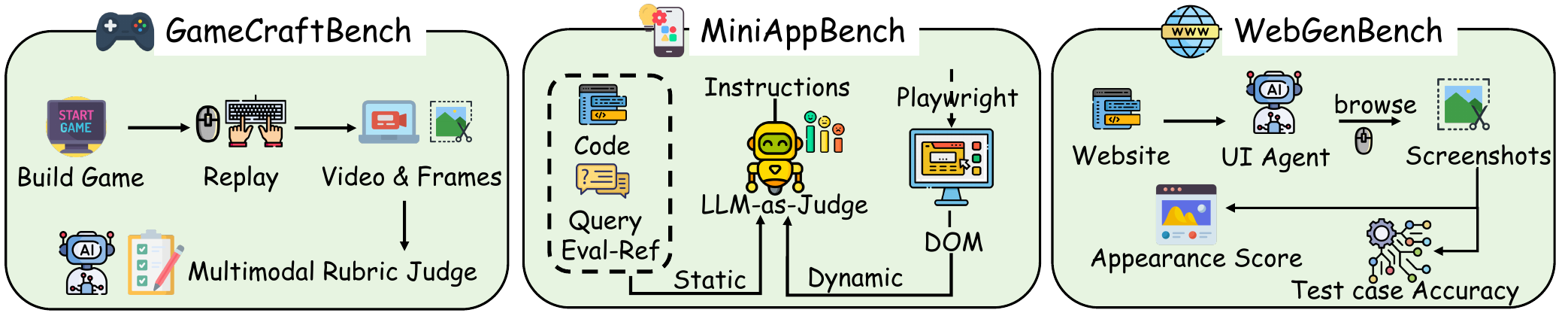}
    \caption{Behavioral evaluation protocols of the three selected source
    benchmarks.} 
    \label{fig:benchmark_evaluation}
    \vspace{-2mm}
\end{figure}

Based on these criteria, we choose \textit{\textbf{GameCraft-Bench}}~\cite{luo2026gamecraftbenchagentsbuildplayable}, \textit{\textbf{MiniAppBench}}~\cite{zhang2026miniappbench}, and \textit{\textbf{WebGen-Bench}}~\cite{lu2026webgenbench}, covering game, application, and web development, respectively.
These benchmarks span diverse software domains and provide high-quality repository-level tasks.
As shown in Figure~\ref{fig:benchmark_evaluation}, they also adopt behavioral evaluation protocols.
Specifically, GameCraft-Bench evaluates recorded gameplay videos and
frames using a multimodal rubric-based judge; MiniAppBench combines
static code analysis with dynamic browser-based evidence collected through
Playwright; and WebGen-Bench employs a UI agent to interact with the
generated website and evaluates both test-case accuracy and visual appearance.
These protocols assess generated software through observable behavior,
allowing us to reliably measure the impact of clarification on subsequent phases.

\textbf{Stage II: Requirement Reconstruction and Intent Abstraction.}
Different benchmarks provide diverse artifacts, including project descriptions, evaluation rubrics, reference code, test cases, etc. 
To derive a unified requirement representation, we adopt the requirement document as our target format, because structured requirement documents are widely established artifacts in real-world development practices~\cite{Fu,zhou2026steeringllmsscalableinteractive}. We leverage an LLM to reconstruct structured requirement documents from these artifacts. This process aims to collect implicit requirements distributed across different artifacts and represent them in an explicit requirement format.

To improve the reliability of the reconstructed requirements, we further introduce a requirement–artifact traceability mechanism. For each reconstructed requirement, the LLM is required to associate it with the corresponding artifact and record the specific evidence source. This mechanism helps prevent the LLM from introducing unsupported requirements or hallucinated details during reconstruction, while providing explicit evidence for subsequent human validation. 
After obtaining the detailed requirement documents, we use an LLM to perform intent abstraction. By removing implementation details and preserving core software objectives, we derive high-level user intents that simulate initial user requests in real-world software development scenarios. The high-level user intents serve as the starting point for requirement clarification.

\textbf{Stage III: Human validation.}
To ensure the reliability of the constructed benchmark, we further conduct human validation. The validation is conducted by two authors. Guided by the requirement quality criteria in ISO/IEC/IEEE 29148:2018~\cite{8559686}, the reconstructed requirement documents are examined for Correctness, Unambiguity, Completeness, and Consistency. Specifically, \textit{\textbf{Correctness}} assesses whether the requirements accurately capture the intended software functionality and behavior. \textit{\textbf{Unambiguity}} checks whether each requirement is clearly stated. \textit{\textbf{Completeness}} examines whether essential functions, constraints, and behaviors are sufficiently covered. \textit{\textbf{Consistency}} checks for conflicts across requirements. Requirement–artifact traceability facilitates human validation by allowing annotators to inspect the supporting evidence for each requirement. Requirements found to be incorrect, ambiguous, incomplete, or inconsistent are regenerated and then re-validated. After validating the reconstructed requirement documents, we further examine the abstraction validity of the high-level user intents, ensuring that they preserve the core software objectives. Only samples that pass both validation stages are included in \benchname{}.

\subsection{Benchmark Statistics.}
Table~\ref{tab:benchmark_statistics} presents the statistics of \benchname{} across its three source benchmarks. In total, \benchname{} contains 341 tasks, comprising 100 tasks from MiniAppBench, 140 from GameCraft-Bench, and 101 from WebGen-Bench. Each task is paired with a human-validated detailed requirement document and its corresponding high-level user intent.
Each detailed requirement document contains 193 atomic requirements and 2,380 tokens on average. In contrast, the corresponding abstracted user intents contain only 13 tokens on average. Notably, this obvious length gap is consistent across all three sources,
indicating that intent abstraction removes many details and leaves substantial room for subsequent requirement clarification.


\begin{table*}[t]
\centering
\caption{Statistics of \benchname{} across its three source benchmarks, including the number of tasks, the average number of atomic requirements per task, and the average token lengths of detailed requirement documents and high-level user intents. ``\#'' denotes the number of items.}
\vspace{-1.5mm}
\label{tab:benchmark_statistics}
\small
\setlength{\tabcolsep}{5pt}
\renewcommand{\arraystretch}{1.08}
\begin{tabular}{l c c c c}
\toprule
\textbf{Source Benchmark}
& \textbf{\# Tasks}
& \textbf{Avg. \# Atomic Req.}
& \textbf{Avg. Doc Tokens}
& \textbf{Avg. Intent Tokens}
\\
\midrule
MiniAppBench~\cite{zhang2026miniappbench}
& 100
& 157
& 1,781
& 10
\\
GameCraft-Bench~\cite{luo2026gamecraftbenchagentsbuildplayable}
& 140
& 240
& 3,526
& 12
\\
WebGen-Bench~\cite{lu2026webgenbench}
& 101
& 165
& 1,383
& 17
\\
\midrule
\rowcolor{gray!10}
\textbf{\benchname{}}
& \textbf{341}
& \textbf{193}
& \textbf{2,380}
& \textbf{13}
\\
\bottomrule
\end{tabular}
\vspace{-2mm}
\end{table*}

\section{Experiment Design}\label{sec:evaluation_setup}

\subsection{Interactive Requirement Clarification}
\begin{figure}
    \centering
    \includegraphics[width=1\linewidth]{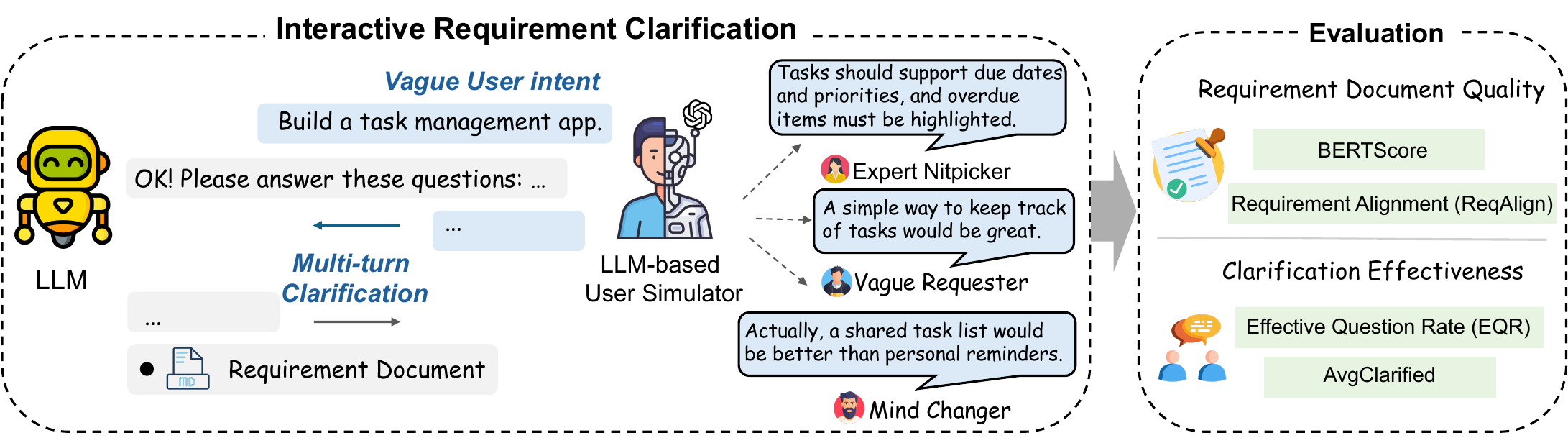}
    \caption{Interactive requirement clarification and evaluation in \benchname{}.}
    \label{fig:experiment_design}
    \vspace{-2mm}
\end{figure}

As illustrated in Figure~\ref{fig:experiment_design}, we formulate requirement clarification as a multi-turn interaction between an LLM and a user. For each \benchname{} task, the LLM initially receives only the user's high-level intent and proactively asks clarification questions. Based on the interaction history, it decides what to clarify next and when to stop asking questions. It then produces a detailed requirement document based on the information gathered during the interaction. The complete interaction, including the clarification questions, user responses, and the final requirement document, is recorded automatically. 
These records are used to evaluate the LLM's ability to produce a complete requirement document through interactive clarification.

\subsection{LLM-based User Simulator}
\label{sec:user_simulator}
Following prior work~\cite{vijayvargiya2026ambigswe,zhou2026tomswe,jin2026reqelicitgym,mu2024clarifygpt}, we employ an LLM to act as the user to engage in multi-turn requirement clarification with the evaluated LLM. Compared with involving human users in interactions, an LLM-based simulator enables large-scale reproducible experiments with substantially lower effort and time costs, while approximating basic user behaviors and producing natural responses under controlled settings~\cite{zhou2025haicosystemecosystemsandboxingsafety,xu2025theagentcompanybenchmarkingllmagents,ni-etal-2026-survey,dou-etal-2025-simulatorarena}. For each task, the LLM-based user simulator is provided with a human-validated detailed requirement document as the ground-truth requirements and answers the clarification questions accordingly.

However, ensuring that the LLM-based user simulator’s responses are consistent with the ground-truth requirements is insufficient to simulate realistic user interactions. Real users differ in how they express their needs, disclose information, and revise their requirements~\cite{BURNAY201461,laporti_athena_2009}. To simulate realistic interactions, we introduce representative user personas based on real-world interaction data. Specifically, we build on SWE-chat~\cite{baumann2026swechatcodingagentinteractions}, a large-scale dataset of multi-turn interactions between real software developers and coding agents. SWE-chat categorizes users into four types: \textbf{\textit{Expert Nitpicker, Vague Requester, Mind Changer}}, and Other. We adopt the first three personas, which together account for approximately 80\% of SWE-chat. To better capture their distinct interaction patterns, we further refine each persona based on corresponding real-world interaction trajectories. \textit{Expert Nitpickers} provide accurate and detailed responses while keeping their goals stable. \textit{Vague Requesters} describe their needs in broad terms and leave many details unspecified. \textit{Mind Changers} revise their requirements as the interaction progresses. These personas are then used to guide the LLM-based user simulator to respond throughout the clarification process, producing more diverse and realistic user interactions.

To assess whether the user simulator exhibits distinct behavioral patterns corresponding to the intended personas, we conduct a \textbf{\textit{blind human evaluation}} with two annotators. Specifically, the annotators are unaware of the specific persona assigned and independently classify each complete interaction trajectory as \textit{Expert Nitpicker}, \textit{Vague Requester}, \textit{Mind Changer}, or \textit{Other}. We randomly sample 60 tasks and obtain interaction trajectories for all three personas, resulting in 180 cases. Following ~\cite{castricato-etal-2025-persona,jin2026reqelicitgym}, we report the average classification accuracy across annotators relative to the assigned personas, and Cohen’s $\kappa$~\cite{kohen1960coefficient}  for inter-annotator agreement. The resulting accuracy is 95.8\% and Cohen’s $\kappa$ is 0.892, suggesting that the user simulator exhibits distinct behavioral patterns corresponding to the intended personas.

\subsection{Evaluation Metrics}
To comprehensively evaluate the performance of LLMs in interactive requirement clarification, we focus on two complementary dimensions:  how effectively LLMs ask clarification questions and requirement document quality.
The former focuses on the effectiveness of the interaction process, whereas the latter focuses on the final outcome of clarification.

\subsubsection{Clarification Effectiveness.}
We evaluate how effectively an LLM obtains new requirement information through interactive clarification using the following two metrics:

\textbf{Effective Question Rate (EQR)} measures the proportion of questions that clarify at least one ground-truth atomic requirement:
\begin{equation}
\mathrm{EQR}
=
\frac{1}{T}
\sum_{t=1}^{T}
\mathbb{I}(c_t > 0),
\end{equation}
where $T$ is the total number of clarification questions and $\mathbb{I}(\cdot)$ is the indicator function. $c_t$ denotes the number of atomic requirements clarified in the $t$-th question-response pair.

\textbf{AvgClarified} measures the average number of atomic requirements clarified per question:
\begin{equation}
\mathrm{AvgClarified}
=
\frac{1}{T}
\sum_{t=1}^{T} c_t.
\end{equation}

\subsubsection{Requirement Document Quality.}
The preceding two metrics evaluate the effectiveness of the interactive clarification process, but do not assess how well the LLM translates the information obtained through clarification into an accurate and complete requirement document. We therefore use two complementary metrics to evaluate requirement document quality:

\textbf{BERTScore} is a widely used automatic metric for text generation~\cite{zhang2020iclr-bertscore}. It matches tokens in the generated and reference texts using contextualized embeddings and computes semantic precision and recall through bidirectional matching. Following prior work~\cite{10.1109/TSE.2025.3554562}, we report BERTScore F1, which combines semantic precision and recall to measure the overall semantic similarity between the generated and ground-truth requirement documents.


\textbf{Requirement Alignment (ReqAlign)} measures how well the generated document aligns with the ground-truth requirements at the atomic requirement level.
Specifically, we decompose the ground-truth requirement document into a set of atomic requirements $\mathcal{R}=\{r_1,r_2,\ldots,r_N\}$. For each atomic requirement $r_i$, we use an LLM-as-a-Judge to determine whether it is correctly \textit{covered}, \textit{omitted}, or \textit{violated} in the generated requirement document. We assign a score $s_i$ to each atomic requirement as follows:
\begin{equation}
s_i =
\begin{cases}
1, & y_i=\mathrm{Covered},\\
0, & y_i=\mathrm{Omitted},\\
-1, & y_i=\mathrm{Violated},
\end{cases}
\end{equation}
where $y_i$ denotes the label assigned to requirement $r_i$. \textbf{\textit{ReqAlign}} is then computed by averaging the scores over all atomic requirements:
\begin{equation}
\mathrm{ReqAlign}
=
\frac{1}{N}\sum_{i=1}^{N}s_i,
\end{equation}
where $N$ denotes the total number of atomic requirements. ReqAlign ranges from $-1$ to $1$, with a higher score indicating better alignment with the ground-truth requirements.

\subsection{Studied LLMs}

\begin{table}[t]
\centering
\caption{Studied Large Language Models}
\label{tab:studied_llms}

\footnotesize
\setlength{\tabcolsep}{4.5pt}
\renewcommand{\arraystretch}{1.10}

\begin{tabular}{@{} l l l c c c @{}}
\toprule

\textbf{Type}
& \textbf{Model}
& \textbf{Publisher}
& \textbf{Size}
& \textbf{Time}
& \textbf{Window Length} \\

\midrule

\multirow{3}{*}{Proprietary}
& GPT-5.6-Sol~\cite{openai2026gpt56}
& OpenAI
& --
& 2026-07
& 1.05M \\

& Claude-Opus-5~\cite{anthropic2026claudeopus5}
& Anthropic
& --
& 2026-07
& 1M \\

& Gemini-3.1-Pro~\cite{googledeepmind2026gemini31pro}
& Google
& --
& 2026-02
& 1M \\

\midrule

\multirow{3}{*}{Open Weight}

& GLM-5.3~\cite{glm5team2026glm5}
& Z.ai
& 753B
& 2026-08
& 1M \\

& DeepSeek-V4-Pro~\cite{deepseekai2026deepseekv4}
& DeepSeek
& 1.6T
& 2026-04
& 1M \\

& Qwen3-235B-A22B~\cite{yang2025qwen3}
& Alibaba
& 235B
& 2025-04
& 128K \\

\bottomrule
\end{tabular}
\vspace{-2mm}
\end{table}
We evaluate six representative LLMs on \benchname{}, including three proprietary models (GPT-5.6-Sol~\cite{openai2026gpt56}, Claude Opus 5~\cite{anthropic2026claudeopus5}, and Gemini-3.1-Pro~\cite{googledeepmind2026gemini31pro}) and three open-weight models (GLM-5.3~\cite{glm5team2026glm5}, DeepSeek-V4-Pro~\cite{deepseekai2026deepseekv4} and Qwen3-235B-A22B~\cite{yang2025qwen3}). Table~\ref{tab:studied_llms} summarizes their basic information. To ensure a fair comparison, we vary only the evaluated LLM while keeping all other experimental settings fixed throughout all experiments, including the LLM-based user simulator, task inputs, persona settings, and interaction protocol.

\subsection{Implementation Details}
We use GPT-5.6-Sol as the foundation model of user simulator and set its temperature to 0.7 to allow natural variation in simulated responses following prior works~\cite{jin2026reqelicitgym,nakash-etal-2026-efficient}. For all evaluated LLMs, we set the temperature to 0.0 to minimize sampling randomness and ensure a fair and stable comparison across models. Similarly, all LLM-as-a-Judge evaluations are conducted with a temperature of 0.0 to improve the consistency of automatic judgments, using DeepSeek-V4.1-Flash~\cite{deepseekv41flash}. 
Following  prior works~\cite{jin2026reqelicitgym,srinivasan2026accordevaluatinguseragentcollaboration}, each requirement clarification session is limited to a maximum of 20 interaction rounds to provide sufficient room for clarification while preventing excessively long interactions.
We automatically record the complete interaction history for subsequent analysis and validation. All experimental configurations and parameters will be released together with our source code to facilitate reproducibility and future benchmarking.

The choices of models for both the LLM-based user simulator and LLM-as-a-Judge are empirically validated. As described in Section~\ref{sec:user_simulator}, human annotators achieve an average persona classification accuracy of 95.8\%, with an inter-annotator Cohen’s $\kappa$ of 0.892, indicating that the GPT-5.6-Sol-based user simulator exhibits distinct behavioral patterns corresponding to the intended personas. For LLM-as-a-Judge, we compare GPT-5.6-Sol, DeepSeek-V4-Pro, and DeepSeek-V4.1-Flash on 200 manually annotated cases. Their judgments achieve accuracies of 92.5\%, 90.0\%, and 91.5\%, respectively, showing comparable performance on this evaluation task. 
Given its competitive accuracy and lower inference cost, we select DeepSeek-V4.1-Flash as the LLM-as-a-Judge.

\section{Evaluation}\label{sec:evaluation}
We conduct a comprehensive empirical study on \benchname{} and mainly investigate the following five research questions:

\begin{itemize}[leftmargin=*, itemsep=2pt, topsep=2pt]
    \item \textbf{RQ1: Overall performance.} We evaluate the overall requirement clarification performance of the studied LLMs on \benchname{} across three representative user personas.
    \item \textbf{RQ2: LLMs’ clarification behavior.} We analyze how LLMs differ in stopping behavior and clarification dynamics across interaction rounds under different user personas.

    \item \textbf{RQ3: Impact of Clarification on Requirement Document Quality.} We 
    quantify the impact of requirement clarification on requirement document quality by comparing documents generated with and without interactive requirement clarification.

    \item \textbf{RQ4: Effect of coding-agent harness.} To assess how coding-agent harnesses affect requirement clarification, we compare standalone LLMs with the same LLMs running within coding-agent harness, while keeping other experimental conditions fixed.

    \item \textbf{RQ5: Impact on Code Generation.} We investigate the impact of clarification and document generation on code implementation, based on the initial high-level intent, directly generated requirement documents, and interactively clarified requirement documents, respectively. 

\end{itemize}

\subsection{RQ1: Overall Performance}

\begin{table*}[t]
\centering
\caption{Overall performance of studied LLMs on \benchname{}.}
\label{tab:overall_performance}

\normalsize
\setlength{\tabcolsep}{2pt}
\renewcommand{\arraystretch}{1.10}

\begin{adjustbox}{max width=\textwidth}
\begin{tabular}{
    l |
    c c c >{\columncolor{gray!20}}c |
    c c c >{\columncolor{gray!20}}c |
    c c c >{\columncolor{gray!20}}c |
    c c c >{\columncolor{gray!20}}c
}
\toprule

\multirow{2}{*}{\textbf{Model}}
& \multicolumn{8}{c|}{\textbf{Clarification Effectiveness}}
& \multicolumn{8}{c}{\textbf{Requirement Document Quality}}
\\

\cmidrule(lr){2-9}
\cmidrule(lr){10-17}

& \multicolumn{4}{c|}{\textbf{EQR}}
& \multicolumn{4}{c|}{\textbf{AvgClarified}}
& \multicolumn{4}{c|}{\textbf{BERTScore F1}}
& \multicolumn{4}{c}{\textbf{ReqAlign}}
\\

\cmidrule(lr){2-5}
\cmidrule(lr){6-9}
\cmidrule(lr){10-13}
\cmidrule(lr){14-17}

& EN. & VR. & MC. & \cellcolor{gray!20}\textbf{Avg.}
& EN. & VR. & MC. & \cellcolor{gray!20}\textbf{Avg.}
& EN. & VR. & MC. & \cellcolor{gray!20}\textbf{Avg.}
& EN. & VR. & MC. & \cellcolor{gray!20}\textbf{Avg.}
\\

\midrule

GPT-5.6-Sol
& 0.359 & 0.080 & 0.150 & 0.196
& 1.396 & 0.199 & 0.356 & 0.650
& \textbf{0.786} & \textbf{0.768} & \textbf{0.768} & \textbf{0.774}
& \textbf{0.842} & \textbf{0.681} & \textbf{0.659} & \textbf{0.727} \\

Claude Opus 5
& 0.463 & 0.153 & 0.216 & 0.277
& 2.013 & 0.432 & 0.576 & 1.007
& \underline{0.743} & 0.705 & 0.707 & 0.718
& \underline{0.720} & \underline{0.504} & \underline{0.507} & \underline{0.577} \\

Gemini-3.1-Pro
& \textbf{0.517} & \textbf{0.255} & \underline{0.229} & \textbf{0.334}
& \textbf{3.510} & \textbf{0.842} & \textbf{0.752} & \textbf{1.702}
& 0.722 & 0.693 & 0.695 & 0.703
& 0.442 & 0.313 & 0.317 & 0.357 \\

\midrule

DeepSeek-V4-Pro
& 0.423 & 0.169 & 0.217 & 0.270
& 2.022 & 0.485 & 0.649 & 1.052
& 0.740 & 0.707 & \underline{0.711} & \underline{0.719}
& 0.610 & 0.434 & 0.450 & 0.498 \\

Qwen3-235B-A22B
& \underline{0.501} & \underline{0.233} & \textbf{0.234} & \underline{0.323}
& \underline{3.331} & \underline{0.841} & \underline{0.744} & \underline{1.639}
& 0.662 & 0.623 & 0.627 & 0.637
& 0.424 & 0.197 & 0.226 & 0.282 \\

GLM-5.3
& 0.425 & 0.169 & 0.203 & 0.266
& 2.033 & 0.497 & 0.585 & 1.039
& 0.737 & \underline{0.708} & 0.707 & 0.717
& 0.618 & 0.471 & 0.474 & 0.521 \\

\bottomrule
\end{tabular}
\end{adjustbox}

\vspace{2pt}
\begin{minipage}{\textwidth}
\footnotesize
* \textbf{EN.}, \textbf{VR.}, and \textbf{MC.} denote the \textit{Expert Nitpicker},
\textit{Vague Requester}, and \textit{Mind Changer} personas, respectively.
\textbf{Avg.} denotes the average performance across the three personas.
\textbf{EQR} (Effective Question Rate) is the proportion of questions that clarify at
least one ground-truth atomic requirement. \textbf{AvgClarified} is the average number
of atomic requirements clarified per question.
\textbf{ReqAlign} (Requirement Alignment) measures how well the generated document
aligns with ground-truth requirements at the atomic-requirement level.
\textbf{Bold} and \underline{underlined} values
indicate the best and second-best results.
\end{minipage}
\vspace{-2mm}
\end{table*}

Table ~\ref{tab:overall_performance} presents the overall performance of the six studied LLMs on \benchname{} across three representative personas. Based on the results, we have the following observations.

\textbf{(1) Clarification effectiveness does not necessarily translate into requirement document quality.}
Different models exhibit inconsistent performance between clarification effectiveness and final requirement document quality.
GPT-5.6-Sol achieves the best requirement document quality, with an average BERTScore F1 of 0.774 and an average ReqAlign of 0.727, outperforming the second-best model (Claude Opus 5) by 26.0\% in ReqAlign.
However, it exhibits the lowest clarification effectiveness, with an average EQR of 0.196 and AvgClarified of 0.650.
In contrast, Gemini-3.1-Pro performs best during the clarification process, achieving the highest average EQR of 0.334 and AvgClarified of 1.702 among all models. However, its average ReqAlign is only 0.357, substantially lower than the 0.727 achieved by GPT-5.6-Sol.
Therefore, clarification effectiveness and requirement document quality reflect two related but distinct capabilities.
Effective clarification helps the model elicit important requirement information, whereas producing a high-quality document further requires the model to correctly understand and integrate the information gathered across multiple turns. This process may further depend on the model’s ability to leverage interaction context, domain knowledge, and reasoning.
\begin{tcolorbox}[
    breakable,
    width=\linewidth,
    colback=gray!15,
    colframe=gray!55,
    boxrule=0pt,
    leftrule=3pt,
    rightrule=0pt,
    toprule=0pt,
    bottomrule=0pt,
    arc=0pt,
    outer arc=0pt,
    boxsep=0pt,
    top=3pt,
    bottom=3pt,
    left=5pt,
    right=4pt
]
\textbf{Finding 1:} Effective clarification does not necessarily lead to high-quality requirement documents. This finding indicates that asking effective clarification questions is only one important part of requirement clarification, while correctly understanding and integrating information across multiple turns is equally critical.
\end{tcolorbox}
\textbf{(2) User personas substantially affect requirement clarification performance.}
All models perform best under the Expert Nitpicker persona, while both requirement document quality and clarification effectiveness decline noticeably when interacting with Vague Requesters and Mind Changers.
For ReqAlign, the average score across the six models is approximately 0.609 under Expert Nitpicker, but decreases to 0.433 and 0.439 under Vague Requester and Mind Changer, respectively. In contrast, BERTScore F1 varies much less, decreasing only from 0.732 under Expert Nitpicker to approximately 0.702 under the other two personas. This suggests that user personas have a particularly pronounced impact on fine-grained requirement alignment.
The differences are even more evident in clarification effectiveness. Among the six models, the average EQR is approximately 0.448 under Expert Nitpicker, compared with only 0.177 and 0.208 under Vague Requester and Mind Changer, respectively. AvgClarified similarly decreases from 2.384 to 0.549 and 0.610.
These results suggest that LLMs remain highly sensitive to user interaction styles. 
They perform substantially better when users provide explicit and detailed responses, but degrade noticeably when user feedback is vague or when requirements change during the interaction.
\begin{tcolorbox}[
    breakable,
    width=\linewidth,
    colback=gray!15,
    colframe=gray!55,
    boxrule=0pt,
    leftrule=3pt,
    rightrule=0pt,
    toprule=0pt,
    bottomrule=0pt,
    arc=0pt,
    outer arc=0pt,
    boxsep=0pt,
    top=3pt,
    bottom=3pt,
    left=5pt,
    right=4pt
]
\textbf{Finding 2:}  User personas have a substantial impact on requirement clarification performance, and current LLMs still exhibit noticeable performance degradation when faced with vague feedback or changing requirements.
\end{tcolorbox}
\textbf{(3) Models exhibit different levels of robustness across user personas.}
In terms of final requirement document quality, GPT-5.6-Sol shows relatively strong robustness to persona changes. Its ReqAlign decreases from 0.842 under Expert Nitpicker to 0.681 and 0.659 under Vague Requester and Mind Changer, retaining approximately 81\% and 78\% of its performance, respectively. GLM-5.3 also shows a relatively stable trend, with ReqAlign decreasing from 0.618 to 0.471 and 0.474, retaining approximately 76\% and 77\%. In contrast, Qwen3-235B-A22B is the most sensitive to persona changes, with ReqAlign dropping from 0.424 under Expert Nitpicker to 0.197 and 0.226, retaining only approximately 46\% and 53\% of its performance. Claude Opus 5, Gemini-3.1-Pro, and DeepSeek-V4-Pro fall between these two extremes. These results indicate that models differ substantially in their ability to maintain requirement understanding and integration when user feedback becomes less informative or requirements evolve during the interaction.
\begin{tcolorbox}[
    breakable,
    width=\linewidth,
    colback=gray!15,
    colframe=gray!55,
    boxrule=0pt,
    leftrule=3pt,
    rightrule=0pt,
    toprule=0pt,
    bottomrule=0pt,
    arc=0pt,
    outer arc=0pt,
    boxsep=0pt,
    top=3pt,
    bottom=3pt,
    left=5pt,
    right=4pt
]
\textbf{Finding 3:}  Models show varying stability across user personas, suggesting that persona robustness is an important dimension for distinguishing their requirement clarification capabilities.
\end{tcolorbox}

\input{data/RQ2_stop}

\subsection{RQ2: LLMs' Clarification Behavior}

To delve into the behavioral patterns of LLMs during interactive requirement clarification, we further analyze their stopping behavior and clarification dynamics across interaction rounds under different user personas.
\subsubsection{Stopping Behavior.}
\label{sec:stopbehave}
Figure ~\ref{fig:rq2-endround-box} illustrates the stopping behavior of the evaluated LLMs under the three personas (i.e., Expert Nitpicker, Mind Changer and Vague Requester). For each task, we record the number of interaction rounds and visualize the distributions using box plots. The left and right boundaries of each box correspond to the first quartile (Q1) and third quartile (Q3), respectively, while the vertical line inside the box denotes the median. The box covers the middle 50\% of the distribution and the diamonds denote the mean number of interaction rounds. These distributions allow us to compare how stopping behavior varies across both personas and models.

\textbf{Overall, the stopping rounds exhibit a relatively consistent trend across the three personas.} Interactions with the Expert Nitpicker tend to end earliest, followed by the Mind Changer, whereas interactions with the Vague Requester generally continue for more rounds. This suggests that user interaction style affects more than the quality of the requirement document and clarification effectiveness. It also influences when a model considers the requirements sufficiently clarified and terminates the interaction. 

\textbf{However, the extent of variation across personas differs among LLMs.} This variation is particularly pronounced for GPT-5.6-Sol and Claude-Opus-5. GPT-5.6-Sol typically terminates after only about 2--3 clarification rounds under the Expert Nitpicker, but increases to around 6 rounds under the Mind Changer and 14--15 rounds under the Vague Requester, with some interactions approaching the maximum round limit. Claude-Opus-5 exhibits a similarly pronounced pattern. DeepSeek-V4-Pro, GLM-5.3, and Qwen3-235B-A22B also exhibit the same general ordering across personas, but with more moderate differences. In contrast, Gemini-3.1-Pro remains concentrated at around 2--3 rounds across all three personas, showing the least variation. Overall, these results show that LLMs differ not only in their general stopping tendencies but also in how strongly they adjust the duration of clarification across user interaction styles.

\definecolor{pENcol}{RGB}{24,63,124}   
\definecolor{pVRcol}{RGB}{138,26,60}   
\definecolor{pMCcol}{RGB}{183,98,11}  

\pgfplotsset{
    rtaxis/.style={
        scale only axis=true,
        width=2.50cm, height=2.35cm,
        xmin=0.4, ymin=0,
        ymajorgrids=true, xmajorgrids=true,
        grid style={draw=black!14, line width=0.3pt},
        axis line style={black, line width=0.6pt},
        tick style={black, line width=0.5pt},
        tick label style={font=\footnotesize},
        label style={font=\small\bfseries},
        title style={font=\small\bfseries, yshift=-3pt,
            text height=1.6ex, text depth=0.5ex},
        ylabel style={yshift=-4pt},
        xlabel style={yshift=2pt},
        legend style={
            draw=none, font=\small, legend columns=-1,
            /tikz/every even column/.append style={column sep=6mm},
        },
        legend image post style={scale=1.5},
    },
}
\tikzset{
    rtbase/.style={line width=0.8pt, mark size=1.3pt,
        mark options={solid, line width=0.45pt}},
    pEN/.style={rtbase, color=pENcol, mark=*},
    pVR/.style={rtbase, color=pVRcol, mark=square*},
    pMC/.style={rtbase, color=pMCcol, mark=triangle*, mark size=1.7pt},
}

\begin{figure*}[t]
\centering
\pgfplotslegendfromname{rtlegend}\\[2pt]
\begin{adjustbox}{max width=\textwidth}
\begin{tabular}{@{}c@{}c@{}c@{}c@{}c@{}c@{}}

\begin{tikzpicture}
\begin{axis}[rtaxis,
    ymax=0.65,
    ytick={0,0.2,0.4,0.6},
    xmax=11.4,
    xtick={1,4,7,11},
    legend to name=rtlegend,
    ylabel={EQR},
    xticklabels={},
    title=\textit{Claude-Opus-5}
]
\addplot[pEN] coordinates {(1,0.402377) (2,0.525298) (3,0.586718)};
\addlegendentry{\textcolor{pENcol}{\bfseries Expert Nitpicker (EN.)}}
\addplot[pVR] coordinates {(1,0.154804) (2,0.185766) (3,0.161470) (4,0.168693) (5,0.169326) (6,0.171334) (7,0.150065) (8,0.163973) (9,0.138581) (10,0.175143) (11,0.124558)};
\addlegendentry{\textcolor{pVRcol}{\bfseries Vague Requester (VR.)}}
\addplot[pMC] coordinates {(1,0.097411) (2,0.191390) (3,0.304049) (4,0.299900) (5,0.310216) (6,0.325665)};
\addlegendentry{\textcolor{pMCcol}{\bfseries Mind Changer (MC.)}}
\end{axis}
\end{tikzpicture}%
&
\begin{tikzpicture}
\begin{axis}[rtaxis,
    ymax=0.65,
    ytick={0,0.2,0.4,0.6},
    xmax=7.4,
    xtick={1,3,5,7},
    yticklabels={},
    xticklabels={},
    title=\textit{DeepSeek-V4-Pro}
]
\addplot[pEN] coordinates {(1,0.397862) (2,0.464686) (3,0.422020)};
\addplot[pVR] coordinates {(1,0.204611) (2,0.190616) (3,0.184034) (4,0.174720) (5,0.145101) (6,0.188435) (7,0.130859)};
\addplot[pMC] coordinates {(1,0.113155) (2,0.236322) (3,0.286868) (4,0.282303) (5,0.282466)};
\end{axis}
\end{tikzpicture}%
&
\begin{tikzpicture}
\begin{axis}[rtaxis,
    ymax=0.65,
    ytick={0,0.2,0.4,0.6},
    xmax=3.4,
    xtick={1,2,3},
    yticklabels={},
    xticklabels={},
    title=\textit{Gemini-3.1-Pro}
]
\addplot[pEN] coordinates {(1,0.489690) (2,0.612039)};
\addplot[pVR] coordinates {(1,0.228873) (2,0.285338) (3,0.264312)};
\addplot[pMC] coordinates {(1,0.134284) (2,0.301766) (3,0.414903)};
\end{axis}
\end{tikzpicture}%
&
\begin{tikzpicture}
\begin{axis}[rtaxis,
    ymax=0.65,
    ytick={0,0.2,0.4,0.6},
    xmax=6.4,
    xtick={1,3,6},
    yticklabels={},
    xticklabels={},
    title=\textit{GLM-5.3}
]
\addplot[pEN] coordinates {(1,0.391423) (2,0.489312)};
\addplot[pVR] coordinates {(1,0.156398) (2,0.183367) (3,0.178075) (4,0.167798) (5,0.158037) (6,0.148106)};
\addplot[pMC] coordinates {(1,0.096816) (2,0.195616) (3,0.318309) (4,0.268247)};
\end{axis}
\end{tikzpicture}%
&
\begin{tikzpicture}
\begin{axis}[rtaxis,
    ymax=0.65,
    ytick={0,0.2,0.4,0.6},
    xmax=18.4,
    xtick={1,6,11,18},
    yticklabels={},
    xticklabels={},
    title=\textit{GPT-5.6-Sol}
]
\addplot[pEN] coordinates {(1,0.306690) (2,0.436030) (3,0.398898)};
\addplot[pVR] coordinates {(1,0.093035) (2,0.091467) (3,0.089487) (4,0.093086) (5,0.077386) (6,0.084714) (7,0.081671) (8,0.078396) (9,0.084543) (10,0.079743) (11,0.071279) (12,0.072653) (13,0.073180) (14,0.077435) (15,0.084923) (16,0.081721) (17,0.072382) (18,0.088560)};
\addplot[pMC] coordinates {(1,0.062498) (2,0.148095) (3,0.190693) (4,0.191403) (5,0.174287) (6,0.185996) (7,0.187548) (8,0.175591)};
\end{axis}
\end{tikzpicture}%
&
\begin{tikzpicture}
\begin{axis}[rtaxis,
    ymax=0.65,
    ytick={0,0.2,0.4,0.6},
    xmax=6.4,
    xtick={1,3,6},
    yticklabels={},
    xticklabels={},
    title=\textit{Qwen3-235B-A22B}
]
\addplot[pEN] coordinates {(1,0.487503) (2,0.527803) (3,0.590660) (4,0.552055)};
\addplot[pVR] coordinates {(1,0.269568) (2,0.272324) (3,0.238909) (4,0.200652) (5,0.217897) (6,0.198538)};
\addplot[pMC] coordinates {(1,0.123042) (2,0.253224) (3,0.373739) (4,0.341727) (5,0.295575)};
\end{axis}
\end{tikzpicture}%
\\[-3pt]

\begin{tikzpicture}
\begin{axis}[rtaxis,
    ymax=4,
    ytick={0,2,4},
    xmax=11.4,
    xtick={1,4,7,11},
    ylabel={AvgClarified},
    xlabel={Round}
]
\addplot[pEN] coordinates {(1,2.530649) (2,2.290053) (3,2.594147)};
\addplot[pVR] coordinates {(1,0.457908) (2,0.637611) (3,0.536811) (4,0.621484) (5,0.514534) (6,0.638714) (7,0.528300) (8,0.543527) (9,0.592386) (10,0.589886) (11,0.601335)};
\addplot[pMC] coordinates {(1,0.218370) (2,0.613784) (3,0.935046) (4,0.976545) (5,1.303750) (6,1.187550)};
\end{axis}
\end{tikzpicture}%
&
\begin{tikzpicture}
\begin{axis}[rtaxis,
    ymax=4,
    ytick={0,2,4},
    xmax=7.4,
    xtick={1,3,5,7},
    yticklabels={},
    xlabel={Round}
]
\addplot[pEN] coordinates {(1,3.232021) (2,2.226255) (3,1.748408)};
\addplot[pVR] coordinates {(1,0.684774) (2,0.701842) (3,0.599921) (4,0.572527) (5,0.497480) (6,0.657161) (7,0.400471)};
\addplot[pMC] coordinates {(1,0.364989) (2,0.889230) (3,0.948827) (4,0.941275) (5,0.955167)};
\end{axis}
\end{tikzpicture}%
&
\begin{tikzpicture}
\begin{axis}[rtaxis,
    ymax=4,
    ytick={0,2,4},
    xmax=3.4,
    xtick={1,2,3},
    yticklabels={},
    xlabel={Round}
]
\addplot[pEN] coordinates {(1,3.670781) (2,3.578562)};
\addplot[pVR] coordinates {(1,0.795156) (2,0.998508) (3,0.830345)};
\addplot[pMC] coordinates {(1,0.352450) (2,1.238611) (3,1.489416)};
\end{axis}
\end{tikzpicture}%
&
\begin{tikzpicture}
\begin{axis}[rtaxis,
    ymax=4,
    ytick={0,2,4},
    xmax=6.4,
    xtick={1,3,6},
    yticklabels={},
    xlabel={Round}
]
\addplot[pEN] coordinates {(1,2.476378) (2,2.117834)};
\addplot[pVR] coordinates {(1,0.521268) (2,0.595966) (3,0.591476) (4,0.471328) (5,0.468295) (6,0.380086)};
\addplot[pMC] coordinates {(1,0.241011) (2,0.660174) (3,0.991542) (4,0.954621)};
\end{axis}
\end{tikzpicture}%
&
\begin{tikzpicture}
\begin{axis}[rtaxis,
    ymax=4,
    ytick={0,2,4},
    xmax=18.4,
    xtick={1,6,11,18},
    yticklabels={},
    xlabel={Round}
]
\addplot[pEN] coordinates {(1,1.869127) (2,1.908027) (3,1.584007)};
\addplot[pVR] coordinates {(1,0.271462) (2,0.325498) (3,0.337500) (4,0.291635) (5,0.241411) (6,0.322670) (7,0.251349) (8,0.309218) (9,0.284442) (10,0.228721) (11,0.243524) (12,0.249730) (13,0.267940) (14,0.285577) (15,0.310589) (16,0.298598) (17,0.192376) (18,0.340288)};
\addplot[pMC] coordinates {(1,0.150129) (2,0.471297) (3,0.600660) (4,0.687424) (5,0.589033) (6,0.586174) (7,0.700436) (8,0.511728)};
\end{axis}
\end{tikzpicture}%
&
\begin{tikzpicture}
\begin{axis}[rtaxis,
    ymax=4,
    ytick={0,2,4},
    xmax=6.4,
    xtick={1,3,6},
    yticklabels={},
    xlabel={Round}
]
\addplot[pEN] coordinates {(1,3.9) (2,3.452712) (3,3.052904) (4,2.629176)};
\addplot[pVR] coordinates {(1,1.300086) (2,1.063289) (3,0.837215) (4,0.650101) (5,0.723188) (6,0.635997)};
\addplot[pMC] coordinates {(1,0.364663) (2,1.024497) (3,1.453120) (4,1.122598) (5,0.927068)};
\end{axis}
\end{tikzpicture}%

\end{tabular}
\end{adjustbox}

\caption{Clarification dynamics across interaction rounds for different models and user personas. For each model--persona pair, EQR and AvgClarified are reported only up to the third quartile (Q3) of the stopping-round distribution to avoid sparsely populated tail rounds.}
\label{fig:rq2-dynamics}
\end{figure*}
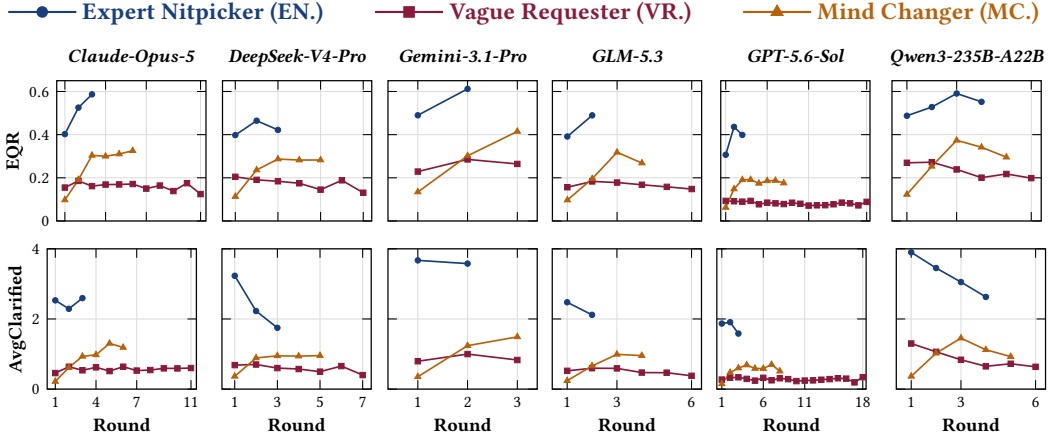

\begin{tcolorbox}[
    breakable,
    width=\linewidth,
    colback=gray!15,
    colframe=gray!55,
    boxrule=0pt,
    leftrule=3pt,
    rightrule=0pt,
    toprule=0pt,
    bottomrule=0pt,
    arc=0pt,
    outer arc=0pt,
    boxsep=0pt,
    top=3pt,
    bottom=3pt,
    left=5pt,
    right=4pt
]

\textbf{Finding 4: }Across models, interactions tend to end after the fewest turns with the Expert Nitpicker and the most turns with the Vague Requester, although the size of this difference varies substantially by model.
\end{tcolorbox}


\subsubsection{Clarification Dynamics.}

As shown in Figure~\ref{fig:rq2-dynamics}, we further analyze how clarification effectiveness evolves across interaction rounds for different models and user personas. Unlike Section~\ref{sec:stopbehave}, which focuses on when models terminate clarification, this analysis investigates how clarification effectiveness changes round by round before termination. Specifically, we compute EQR and AvgClarified at each interaction round. Because tasks terminate at different rounds, the number of active tasks decreases as the interaction progresses, making statistics at very late rounds less representative.
To mitigate this issue, for each model--persona pair, we report round-wise results only up to the third quartile (Q3) of its stopping-round distribution. This avoids drawing conclusions from sparsely populated tail rounds while retaining all tasks in the analysis.

\textbf{The six evaluated models generally exhibit similar trends across the three personas.}
For the Expert Nitpicker, most requirement information is obtained within the first few rounds. EQR and AvgClarified are relatively high early in the interaction, and the models typically stop after only a few rounds. This can be characterized as a short but information-intensive clarification process. In contrast, interactions with the Mind Changer tend to last longer, with EQR and AvgClarified often showing a noticeable increase after several rounds. This suggests that new requirement information is not concentrated at the beginning, but often emerges later as the user revises or supplements the requirements. 
For the Vague Requester, EQR and AvgClarified generally remain low across interaction rounds, and the models tend to continue asking clarification questions for more rounds. This pattern can be characterized as a low-yield but sustained clarification process, in which requirement information accumulates gradually over multiple interactions.


\textbf{Under the Vague Requester, the models broadly exhibit three representative clarification dynamics.}
Gemini-3.1-Pro and Qwen3-235B-A22B achieve relatively high per-round clarification gains, with both EQR and AvgClarified generally higher than those of the other models. Meanwhile, their interactions also terminate relatively early, particularly for Gemini-3.1-Pro, whose stopping-round Q3 is only 3. In contrast, Claude-Opus-5, DeepSeek-V4-Pro, and GLM-5.3 maintain relatively low values on both metrics, exhibiting a comparatively stable low-gain clarification process, with stopping-round Q3 values ranging from approximately 6 to 11.
GPT-5.6-Sol also remains in a low-gain state for much of the interaction, but unlike the other low-gain models, it continues clarification for substantially more rounds, with a stopping-round Q3 of 18.
These results indicate that models differ not only in their ability to obtain useful requirement information within each round, but also in their stopping behavior under sustained low-gain interactions.

\definecolor{bertlight}{RGB}{250,193,184}
\definecolor{bertdark}{RGB}{180,45,85}
\definecolor{reqlight}{RGB}{184,218,238}
\definecolor{reqdark}{RGB}{42,99,172}
\definecolor{berttext}{RGB}{0,0,0}
\definecolor{reqtext}{RGB}{0,0,0}
\def\rqthreeRows{%
    0/0.713616/0.060464/0.322215/0.405253,
    1/0.655854/0.062367/0.224602/0.352481,
    2/0.666046/0.037122/0.147818/0.209466,
    3/0.676816/0.042296/0.210627/0.287497,
    4/0.599396/0.037964/0.082722/0.199587,
    5/0.670811/0.046683/0.283401/0.237611
}

\begin{figure*}[t]
\centering
\begin{tikzpicture}
\begin{axis}[
    width=0.5\textwidth,
    height=0.25\textwidth,
    xmin=-0.5,
    xmax=5.5,
    ymin=0,
    ymax=1,
    ytick={0,0.5,1},
    ymajorgrids=true,
    grid style={draw=black!16, line width=0.35pt},
    axis line style={black, line width=0.75pt},
    tick style={black, line width=0.6pt},
    ylabel={Performance},
    ylabel style={font=\bfseries},
    xtick={0,1,2,3,4,5},
    xticklabels={
        {GPT-5.6-Sol},
        {Claude Opus 5},
        {Gemini-3.1-Pro},
        {DeepSeek-V4-Pro},
        {Qwen3-235B-A22B},
        {GLM-5.3}
    },
    xticklabel style={font=\footnotesize, rotate=18, anchor=north east},
    yticklabel style={font=\footnotesize},
    legend columns=1,
    legend style={
        at={(1,0.5)},
        anchor=west,
        xshift=10pt,
        draw=none,
        fill=none,
        font=\footnotesize,
        cells={anchor=west},
        row sep=1pt,
        /tikz/every even column/.append style={column sep=2pt}
    },
    legend image code/.code={
        \draw[#1] (0cm,-0.1cm) rectangle (0.28cm,0.1cm);
    },
    every axis plot/.append style={draw=black, line width=0.6pt},
    clip=false
]

\foreach \x/\BERTbaseVal/\BERTgainVal/\ReqbaseVal/\ReqgainVal in \rqthreeRows {%
    \pgfmathsetmacro{\BERTtotalVal}{\BERTbaseVal+\BERTgainVal}
    \pgfmathsetmacro{\ReqtotalVal}{\ReqbaseVal+\ReqgainVal}

    \pgfmathsetmacro{\BERTleft}{\x-0.28}
    \pgfmathsetmacro{\BERTright}{\x}
    \pgfmathsetmacro{\Reqleft}{\x}
    \pgfmathsetmacro{\Reqright}{\x+0.28}

    \edef\drawRQbars{%
        \noexpand\path[fill=bertlight, draw=black, line width=0.6pt]
            (axis cs:\BERTleft,0)
            rectangle
            (axis cs:\BERTright,\BERTbaseVal);
        \noexpand\path[fill=bertdark, draw=black, line width=0.6pt]
            (axis cs:\BERTleft,\BERTbaseVal)
            rectangle
            (axis cs:\BERTright,\BERTtotalVal);

        \noexpand\path[fill=reqlight, draw=black, line width=0.6pt]
            (axis cs:\Reqleft,0)
            rectangle
            (axis cs:\Reqright,\ReqbaseVal);
        \noexpand\path[fill=reqdark, draw=black, line width=0.6pt]
            (axis cs:\Reqleft,\ReqbaseVal)
            rectangle
            (axis cs:\Reqright,\ReqtotalVal);
    }%
    \drawRQbars
}%

\addlegendimage{fill=bertlight, draw=black}
\addlegendentry{\textcolor{berttext}{BERTScore F1: w/o Clarification}}

\addlegendimage{fill=bertdark, draw=black}
\addlegendentry{\textcolor{berttext}{BERTScore F1: Clarification Gain}}

\addlegendimage{fill=reqlight, draw=black}
\addlegendentry{\textcolor{berttext}{ReqAlign: w/o Clarification}}

\addlegendimage{fill=reqdark, draw=black}
\addlegendentry{\textcolor{berttext}{ReqAlign: Clarification Gain}}

\end{axis}
\end{tikzpicture}

\label{-2mm}
\caption{Effect of clarification on BERTScore F1 and ReqAlign across different models.}
\label{fig:rq3-interaction}
\vspace{-2mm}
\end{figure*}
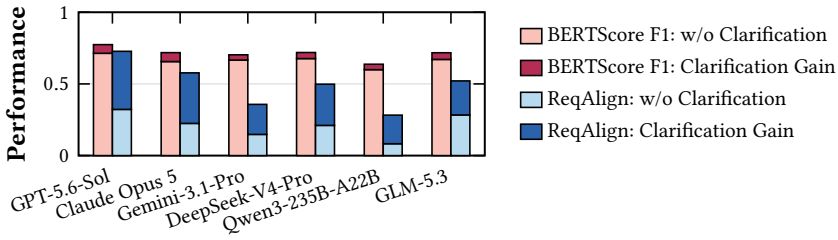

\begin{tcolorbox}[
    breakable,
    width=\linewidth,
    colback=gray!15,
    colframe=gray!55,
    boxrule=0pt,
    leftrule=3pt,
    rightrule=0pt,
    toprule=0pt,
    bottomrule=0pt,
    arc=0pt,
    outer arc=0pt,
    boxsep=0pt,
    top=3pt,
    bottom=3pt,
    left=5pt,
    right=4pt
]
\textbf{Finding 5:} Different personas lead to distinct clarification dynamics that are generally consistent across models. Expert Nitpicker interactions are short but information-intensive, Mind Changer interactions tend to reveal new information later, and Vague Requester interactions accumulate information gradually over more rounds. Under vague requests, models further differ in their per-round information gains and stopping behavior.
\end{tcolorbox}

\subsection{RQ3: Impact of Clarification on Requirement Document Quality}

As shown in Figure~\ref{fig:rq3-interaction}, we investigate the impact of interactive clarification on final requirement document quality by comparing two settings: directly generating requirement documents from initial user intents and generating them after multi-turn requirement clarification.

\textbf{Overall, all models benefit from requirement clarification, producing more complete and clear requirement documents from vague user intents.}
Specifically, the average BERTScore F1 across the six models increases from 0.664 to 0.712, with a 7.2\% relative improvement.
Additionally, the improvement in fine-grained requirement alignment is substantially more pronounced. The average ReqAlign score rises from 0.212 to 0.494 with a relative improvement of 133.1\%. This trend is also consistent across different models, with relative improvements ranging from 83.8\% (GLM-5.3) to 241.3\% (Qwen3-235B-A22B).
The substantially larger improvement in ReqAlign suggests that clarification particularly helps models capture fine-grained requirements that are missing from the initial user intent.
Notably, GPT-5.6-Sol achieves the highest ReqAlign both with and without interactive clarification, demonstrating its consistently strong requirement document generation capability across the two settings.

\begin{tcolorbox}[
    breakable,
    width=\linewidth,
    colback=gray!15,
    colframe=gray!55,
    boxrule=0pt,
    leftrule=3pt,
    rightrule=0pt,
    toprule=0pt,
    bottomrule=0pt,
    arc=0pt,
    outer arc=0pt,
    boxsep=0pt,
    top=3pt,
    bottom=3pt,
    left=5pt,
    right=4pt
]
\textbf{Finding 6:} Requirement clarification consistently improves requirement document quality across all models, with substantially larger gains in fine-grained requirement alignment than in overall semantic similarity. GPT-5.6-Sol further maintains the strongest ReqAlign performance both with and without clarification.
\end{tcolorbox}


\definecolor{claudedark}{RGB}{42,99,172}
\definecolor{claudelight}{RGB}{184,218,238}
\definecolor{claudelegenddark}{RGB}{42,99,172}
\definecolor{claudelegendlight}{RGB}{90,145,185}

\definecolor{gptdark}{RGB}{180,45,85}
\definecolor{gptlight}{RGB}{250,193,184}
\definecolor{gptlegenddark}{RGB}{180,45,85}
\definecolor{gptlegendlight}{RGB}{238,128,120}

\pgfplotstableread{
    Label GPT Codex
    EN. 0.7535 0.7075
    VR. 0.7201 0.6832
    MC. 0.7245 0.6797
    {w/o clari.} 0.6762 0.6568
}\claudeA

\pgfplotstableread{
    Label GPT Codex
    EN. 0.7590 0.6443
    VR. 0.5173 0.4798
    MC. 0.5363 0.4821
    {w/o clari.} 0.2586 0.1391
}\claudeB

\pgfplotstableread{
    Label GPT Codex
    EN. 0.2892 0.3922
    VR. 0.0949 0.1839
    MC. 0.1021 0.1896
}\claudeC

\pgfplotstableread{
    Label GPT Codex
    EN. 1.6075 3.0000
    VR. 0.3611 0.9433
    MC. 0.3733 0.8063
}\claudeD


\pgfplotstableread{
    Label GPT Codex
    EN. 0.7900 0.7753
    VR. 0.7673 0.7464
    MC. 0.7718 0.7535
    {w/o clari.} 0.7133 0.7243
}\gptA

\pgfplotstableread{
    Label GPT Codex
    EN. 0.8492 0.6454
    VR. 0.6635 0.4764
    MC. 0.6673 0.5105
    {w/o clari.} 0.2935 0.3114
}\gptB

\pgfplotstableread{
    Label GPT Codex
    EN. 0.3395 0.3926
    VR. 0.0980 0.1981
    MC. 0.1759 0.1977
}\gptC

\pgfplotstableread{
    Label GPT Codex
    EN. 2.1832 2.9945
    VR. 0.4357 0.8742
    MC. 0.7612 1.0195
}\gptD


\newcommand{\rqfourdepthstrut}{%
    \node[
        font=\small,
        rotate=35,
        anchor=east,
        opacity=0
    ] at (current axis.south) {w/o clari.};%
}

%

\newcommand{\rqfourpanel}[4]{%
\begin{tikzpicture}
    \begin{axis}[
        ybar=0pt,
        bar width=6pt,
        height=3cm,
        width=4.2cm,
        enlarge x limits=0.25,
        ymin=0,
        ymax=#1,
        axis y line*=none,
        axis x line*=none,
        xtick=data,
        xticklabels from table={#2}{Label},
        xticklabel style={
            font=\small,
            rotate=35,
            anchor=east
        },
        tick label style={font=\normalsize},
        every axis plot/.append style={
            draw=black,
            line width=0.5pt
        },
        every axis/.append style={
            line width=0.5pt
        },
    ]
        \addplot[fill=#3]
            table[y=GPT, x expr=\coordindex] {#2};

        \addplot[fill=#4]
            table[y=Codex, x expr=\coordindex] {#2};
    \end{axis}

    \rqfourdepthstrut
\end{tikzpicture}%
}

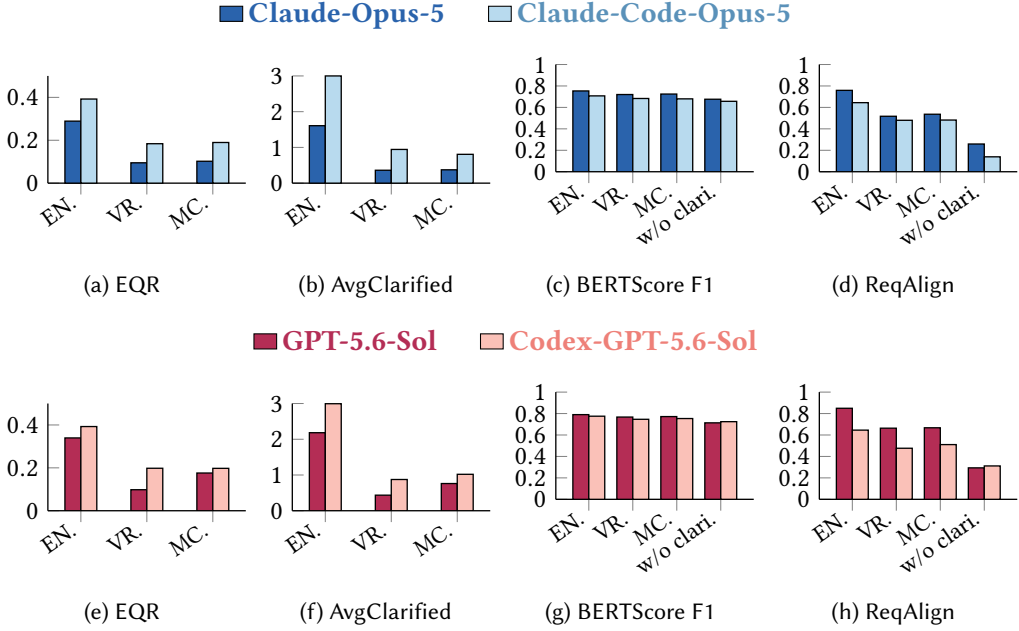
\begin{figure*}[t]
\centering

\pgfplotslegendfromname{rqfourlegendclaude}\\[-7pt]

\subfloat[EQR]{\rqfourpanel{0.5}{\claudeC}{claudedark}{claudelight}}%
\hspace{2mm}%
\subfloat[AvgClarified]{\rqfourpanel{3}{\claudeD}{claudedark}{claudelight}}%
\hspace{2mm}%
\subfloat[BERTScore F1]{\rqfourpanel{1}{\claudeA}{claudedark}{claudelight}}%
\hspace{2mm}%
\subfloat[ReqAlign]{\rqfourpanel{1}{\claudeB}{claudedark}{claudelight}}%

\par\vspace{2mm}

\pgfplotslegendfromname{rqfourlegendgpt}\\[-7pt]

\subfloat[EQR]{\rqfourpanel{0.5}{\gptC}{gptdark}{gptlight}}%
\hspace{2mm}%
\subfloat[AvgClarified]{\rqfourpanel{3}{\gptD}{gptdark}{gptlight}}%
\hspace{2mm}%
\subfloat[BERTScore F1]{\rqfourpanel{1}{\gptA}{gptdark}{gptlight}}%
\hspace{2mm}%
\subfloat[ReqAlign]{\rqfourpanel{1}{\gptB}{gptdark}{gptlight}}%

\par


\makebox[0pt][l]{\raisebox{0pt}[0pt][0pt]{%
\begin{tikzpicture}
\begin{axis}[
    hide axis,
    width=3cm,
    height=2cm,
    xmin=0, xmax=1,
    ymin=0, ymax=1,
    legend columns=2,
    legend to name=rqfourlegendclaude,
    legend image code/.code={
        \draw[#1] (0cm,-0.1cm) rectangle (0.3cm,0.1cm);
    },
    legend style={
        draw=none,
        font=\large,
        /tikz/every even column/.append style={column sep=5mm}
    }
]
\addplot[fill=claudedark, draw=black] coordinates {(0,0)};
\addlegendentry{
    \textcolor{claudelegenddark}{\textbf{Claude-Opus-5}}
}

\addplot[fill=claudelight, draw=black] coordinates {(0,0)};
\addlegendentry{
    \textcolor{claudelegendlight}{\textbf{Claude-Code-Opus-5}}
}
\end{axis}
\end{tikzpicture}%
}}


\makebox[0pt][l]{\raisebox{0pt}[0pt][0pt]{%
\begin{tikzpicture}
\begin{axis}[
    hide axis,
    width=3cm,
    height=2cm,
    xmin=0, xmax=1,
    ymin=0, ymax=1,
    legend columns=2,
    legend to name=rqfourlegendgpt,
    legend image code/.code={
        \draw[#1] (0cm,-0.1cm) rectangle (0.3cm,0.1cm);
    },
    legend style={
        draw=none,
        font=\large,
        /tikz/every even column/.append style={column sep=5mm}
    }
]
\addplot[fill=gptdark, draw=black] coordinates {(0,0)};
\addlegendentry{
    \textcolor{gptlegenddark}{\textbf{GPT-5.6-Sol}}
}

\addplot[fill=gptlight, draw=black] coordinates {(0,0)};
\addlegendentry{
    \textcolor{gptlegendlight}{\textbf{Codex-GPT-5.6-Sol}}
}
\end{axis}
\end{tikzpicture}%
}}

\vspace{-5mm}
\caption{
Comparison of requirement-document quality and clarification effectiveness
between base models and their coding-agent harness counterparts.
\textit{EN.}, \textit{VR.}, and \textit{MC.} denote Expert Nitpicker, Vague Requester, and Mind Changer, respectively.
\textit{w/o clari.} denotes directly generating the requirement document without interactive requirement clarification.
}
\label{fig:rq4_agent_harness}

\end{figure*}

\subsection{RQ4: Effect of Coding-Agent Harness}

As shown in Figure~\ref{fig:rq4_agent_harness}, we further compare Claude-Opus-5 and GPT-5.6-Sol with their corresponding agent-harness counterparts (i.e., Claude Code~\cite{anthropic_claude_code} and Codex~\cite{openai_codex}), in terms of both clarification effectiveness and final requirement-document quality. In practical repository-level software development, developers increasingly interact with coding agents such as Claude Code and Codex rather than directly prompting base LLMs. These agent harnesses introduce additional scaffolding for task execution and software development workflows, which may also affect how models clarify and organize user requirements before implementation.

\textbf{Overall, agent harnesses improve clarification effectiveness, but this improvement does not translate into higher-quality requirement documents.} Across the three personas, Claude Code and Codex consistently achieve higher EQR and AvgClarified than their corresponding base models. This indicates that agent harnesses help models ask more effective clarification questions through each interaction. However, the opposite trend is observed in the final requirement documents. Claude-Opus-5 and GPT-5.6-Sol generally achieve higher BERTScore F1 and ReqAlign than their agent-harness counterparts, with the gap being particularly pronounced for ReqAlign. For example, under the Expert Nitpicker, ReqAlign decreases by 15.2\% for Claude and 24.0\% for GPT when using coding-agent harnesses. The same trend is observed for Vague Requester and Mind Changer.  These results reveal a trade-off introduced by coding-oriented agent harnesses. While they facilitate more effective requirement clarification, their scaffolding may compromise some of the base model's capabilities needed to produce complete and accurate requirement documents.

\begin{tcolorbox}[
    breakable,
    width=\linewidth,
    colback=gray!15,
    colframe=gray!55,
    boxrule=0pt,
    leftrule=3pt,
    rightrule=0pt,
    toprule=0pt,
    bottomrule=0pt,
    arc=0pt,
    outer arc=0pt,
    boxsep=0pt,
    top=3pt,
    bottom=3pt,
    left=5pt,
    right=4pt
]
\textbf{Finding 7: } Coding-oriented agent harnesses improve clarification effectiveness, but this gain is accompanied by lower requirement-document quality than their corresponding base models, suggesting a trade-off between interactive clarification and final document generation.
\end{tcolorbox}

\subsection{RQ5: Impact on Code Generation}
To further investigate the impact of requirement clarification on subsequent code generation, we conduct code-generation experiments on the three source benchmarks of \benchname{}: GameCraft-Bench, MiniAppBench, and WebGen-Bench. We follow the original evaluation protocols of each benchmark.
We use Codex with GPT-5.6-Sol to generate code under three requirement settings: the initial intent, the directly generated requirement document, and the interactively clarified requirement document.
Both types of requirement documents are generated by GPT-5.6-Sol, and scores are averaged across the Expert, Vague, and Mind Changer personas.
We compare performance across these settings to assess the effects of requirement documentation and interactive clarification on code generation.
Table~\ref{tab:benchmark_scores} summarizes the score results across the three benchmarks.

\begin{table}[t]
\centering
\caption{
Code-generation performance across three source benchmarks under three input settings:
\textbf{Intent} (the initial user intent),
\textbf{Direct Doc.} (a requirement document generated directly from the initial intent), and
\textbf{Clarified Doc.} (a requirement document generated after interactive clarification).
}
\label{tab:benchmark_scores}
\begin{tabular}{l|ccc}
\toprule
\multirow{2}{*}{\textbf{Benchmark}}
& \multicolumn{3}{c}{\textbf{Score}} \\
\cmidrule(lr){2-4}
& \textbf{Intent}
& \textbf{Direct Doc.}
& \textbf{Clarified Doc.} \\
\midrule
GameCraft-Bench & 0.012 & \underline{0.273} & \textbf{0.418} \\
MiniAppBench    & 0.451 & \underline{0.543} & \textbf{0.943}  \\
WebGen-Bench   & 0.271 & \underline{0.384} & \textbf{0.659}  \\
\bottomrule
\end{tabular}
\vspace{-2mm}
\end{table}

\textbf{Overall, requirement documentation itself improves code-generation performance, while interactive clarification provides further gains on top of it.} Across all three source benchmarks, code-generation performance consistently improves when moving from the initial intent to a directly generated requirement document and then to an interactively clarified requirement document. Even without any additional user feedback, guiding the LLM to first transform the initial intent into a requirement document and then implement based on that document is more effective than generating code directly from the initial intent. The score increases from 0.012 to 0.273 on GameCraft-Bench, from 0.451 to 0.543 on MiniAppBench, and from 0.271 to 0.384 on WebGen-Bench. Furthermore, interactive clarification leads to additional improvements on all three benchmarks. This consistent trend shows that the value of interactive clarification extends beyond improving requirement-document quality. The requirement information clarified through multi-turn interaction can further translate into tangible gains in subsequent code implementation.

\begin{tcolorbox}[
    breakable,
    width=\linewidth,
    colback=gray!15,
    colframe=gray!55,
    boxrule=0pt,
    leftrule=3pt,
    rightrule=0pt,
    toprule=0pt,
    bottomrule=0pt,
    arc=0pt,
    outer arc=0pt,
    boxsep=0pt,
    top=3pt,
    bottom=3pt,
    left=5pt,
    right=4pt
]
\textbf{Finding 8: } Requirement documentation improves code-generation performance even without additional user feedback, while interactive requirement clarification provides further gains across all three source benchmarks.
\end{tcolorbox}

\section{Discussion}
\subsection{Practical Implications}
\textbf{(1) LLMs should proactively clarify requirements before implementation.} In practice, developers often provide high-level and underspecified requests. Rather than proceeding directly to implementation and filling in missing details based on their own assumptions, LLMs should first identify unclear or missing requirements and actively ask developers for clarification. Our study shows that such interactive clarification can elicit additional requirement information that ultimately improves subsequent code-generation performance. This suggests that requirement clarification should be treated as an explicit capability of LLMs, rather than leaving the burden of fully specifying requirements to developers in the initial request.

\textbf{(2) Requirement documentation is beneficial even without additional user feedback.} We find that guiding the LLM to first transform the initial intent into an explicit requirement document, and then generate code based on that document, consistently outperforms direct code generation from the initial intent. This suggests that a requirement document is not merely a record of clarification outcomes, but also serves as an important intermediate representation between user intent and code implementation. In practice, even when further clarification with the developer is unavailable, introducing an explicit requirement-document generation stage before coding can still improve subsequent implementation compared with directly using the original request.

\textbf{(3) Requirement engineering and code implementation may be better handled by different forms of model systems.} We observe that coding-oriented agent harnesses such as Claude Code and Codex ask more effective clarification questions, yet produce lower-quality final requirement documents than their corresponding base models. This suggests that scaffolding designed for code execution may not be equally well suited to requirement-document generation. A more suitable workflow may therefore decouple the requirement and implementation stages. Developers can first use a base model for requirement analysis, integration of clarified information, and final requirement-document generation, and then pass the resulting document to coding-agent harnesses such as Claude Code or Codex for repository-level implementation. Such a staged workflow can leverage the respective strengths of base models in requirement documentation and coding-agent harnesses in code execution and software-development tooling, rather than relying on a single agent to handle the entire process from underspecified requests to final implementation.

\subsection{Threats to Validity}

\textbf{External Validity. }
A primary threat to external validity is the reliability of user simulation. Real developers vary substantially in how they express requirements, provide information, and revise their needs, and an LLM-based user simulator may not fully capture this diversity. To mitigate this threat, we derive three representative user personas from SWE-chat~\cite{baumann2026swechatcodingagentinteractions} to guide the LLM-based user simulator. SWE-chat is a large-scale dataset of multi-turn conversations between software developers and coding agents, reflecting diverse interaction behaviors in real-world development settings. We also conduct human validation to verify that the user simulator exhibits distinct behavioral patterns corresponding to the intended personas.

\textbf{Internal Validity.}
A potential threat to internal validity arises from possible errors in the interaction protocol, prompt design, model configuration, and experimental execution. Such errors may affect the observed clarification behavior or subsequent code-generation evaluation. To reduce this risk, we use a consistent interaction protocol and standardized prompt templates across all evaluated models and settings, and keep the experimental configurations fixed whenever possible. In addition, we have open-sourced our implementation, experimental data, and all prompt templates, enabling external validation and community feedback for continuous improvement.

\textbf{Construct Validity.}
A primary threat to construct validity is the quality of the ground-truth requirement documents in \benchname{}. To mitigate this threat, two authors manually validate the reconstructed requirement documents by examining their correctness, unambiguity, completeness, and consistency. Requirement–artifact traceability further helps the two validators verify the implementation evidence supporting each requirement. Requirements that fail the validation are regenerated and re-examined. We also validate the abstraction validity of the high-level user intents to ensure that they preserve the core software objectives. Only samples that pass both stages of human validation are included in \benchname{}.

\section{Conclusion}\label{sec:conclusion}
In this paper, we propose \benchname{}, the first interactive requirement clarification benchmark for repository-level software development, which aims to evaluate LLMs’ ability to transform vague user intents into clear and complete requirement documents through multi-turn interactions.
To better simulate realistic interactions, we further employ LLM-based user simulators with three representative personas.
Based on \benchname{}, we conduct a systematic empirical study.
The results reveal that clarification effectiveness does not necessarily lead to high-quality requirement documents, and that model performance varies substantially across different user interaction styles. 
We also find that coding-agent harnesses do not further enhance the quality of derived requirement documents.
These findings offer practical implications for developers using LLMs for software development. 
More broadly, our work highlights that software development agents’ autonomy should encompass understanding user intent through proactive interaction and aligning development goals accordingly. This offers a user-centered perspective on the design and evaluation of future autonomous development systems.

\section*{Data Availability}\label{sec:dataavail}
The code and data are available at \cite{reqclarifybench2026}.


\bibliographystyle{ACM-Reference-Format}
\bibliography{main}


\end{document}